\documentclass{article}
\usepackage{arxiv}
\usepackage[utf8]{inputenc} 
\usepackage[T1]{fontenc}    
\usepackage{hyperref}       
\usepackage{url}         
\usepackage{booktabs}    
\usepackage{amsfonts}      
\usepackage{nicefrac}      
\usepackage{microtype}     
\usepackage{lipsum}
\usepackage{float}
\usepackage{graphicx}
\usepackage{caption}
\usepackage{subcaption}
\usepackage{titlesec}
\usepackage{multirow}
\usepackage{amsmath}
\usepackage{soul,color}
\usepackage{multicol}
\usepackage{placeins} 
\usepackage[nodisplayskipstretch]{setspace}

\title{Composition based crystal materials symmetry prediction using machine learning with enhanced descriptors }

\author{
  Yuxin Li\\
School of Mechanical Engineering\\
  Guizhou University \\
  Guiyang China 550025 \\
 \And
   Rongzhi Dong, Wenhui Yang\\
School of Mechanical Engineering\\
  Guizhou University \\
  Guiyang China 550025 \\
   \And
 Jianjun Hu \thanks{Corresponding author: J.H. (webpage)}\\
  Department of Computer Science and Engineering\\
  University of South Carolina\\
  Columbia, SC, 29201, USA \\
  \texttt{jianjunh@cse.sc.edu} \\
}

\begin{document}
\maketitle

\begin{abstract}
Geometric information such as the space groups and crystal systems plays an important role in the properties of crystal materials. Prediction of crystal system and space group thus has wide applications in crystal material property estimation and structure prediction. Previous works on experimental X-ray diffraction (XRD) and density functional theory (DFT) based structure determination methods achieved outstanding performance, but they are not applicable for large-scale screening of materials compositions. There are also machine learning models using Magpie descriptors for composition based material space group determination, but their prediction accuracy only ranges between 0.638 and 0.907 in different kinds of crystals. Herein, we report an improved machine learning model for predicting the crystal system and space group of materials using only the formula information. Benchmark study on a dataset downloaded from Materials Project Database shows that our random forest models based on our new descriptor set, achieve significant performance improvements compared with previous work with accuracy scores ranging between 0.712 and 0.961 in terms of space group classification. Our model also shows large performance improvement for crystal system prediction. Trained models and source code are freely available at \url{https://github.com/Yuxinya/SG_predict}

\end{abstract}


\section{Introduction}

According to the degree of geometric form symmetry, the crystals can be divided into different crystal systems and space groups\cite{paufler2020fedorov,doi:10.1021/acsomega.1c00781}. Determining the symmetry information, particularly the space group, provides wide applications for crystal material property prediction\cite{xie2018crystal} and crystal structure prediction\cite{po2017symmetry,sato2020adjusting}. 
Recently, we proposed the method of knowledge-rich approach for crystal structure prediction\cite{hu2020contact,hu2020distance,hu2021alphacrystal,yang2021crystal}, which is inspired by the recent advances in protein structure prediction (PSP)\cite{zheng2019deep,senior2020improved} which predicts protein structures using the predicted distance matrix. In our approach, various global optimization algorithms\cite{hu2020contact} such as genetic algorithm\cite{hu2020distance}, neural networks\cite{hu2021alphacrystal}, differential evolution
algorithm\cite{yang2021crystal} have been used to reconstruct the crystal structure atom coordinates. However, it is indispensable to provide the space group information for a given materials composition before predicting their structures.

Several machine learning models have been proposed to predict the space groups of crystals. Several studies
developed machine learning approaches for space group classification by the X-ray diffraction (XRD)\cite{holder2019tutorial} data of materials. 
Suzuki et al.\cite{suzuki2020symmetry} emphasised on demonstrating the potential of simple machine learning techniques suitable for knowledge discovery and real-world experiments. Their tree-ensemble-based machine learning model works with over 90\% accuracy for crystal system classification based on powder X-ray diffraction patterns.
Park et al.\cite{park2017classification} developed three convolutional neural networks (CNNs) for the space group, extinction group and crystal system classification of 150,000 powder XRD patterns, which returned test accuracies of 81.14, 83.83 and 94.99$\%$ respectively.
Vecsei et al.\cite{vecsei2019neural} studied the problem of space group determination from powder X-ray diffraction patterns by using fully connected neural networks and convolutional neural networks and then tested those two models on the orher database. 
Oviedo et al.\cite{oviedo2019fast} proposed a supervised machine learning framework for rapid crystal structure identification of novel materials from thin-film XRD measurements.
Chakraborty et al.\cite{chakraborty2020see} performed augmentation of thin filmed X-ray diffraction patterns and developed a high accuracy model for lattice classification from X-ray diffraction.
Zaloga et al.\cite{zaloga2020crystal} identified crystal systems and symmetry space groups by full-profile X-ray diffraction patterns using convolutional neural networks and explored the factors that affect the classification performance.
Ziletti et al.\cite{ziletti2018insightful} represented crystals by calculating a diffraction image, then constructed a deep learning neural network model for classification which achieves robust performance even in the presence of highly defective structures.

Although the XRD based symmetry prediction algorithms can achieve good performance in space group classification, they have several limitations. The performance of XRD based methods is frequently influenced by low-quality X-ray diffraction data\cite{wang2010crystal}. Moreover, it is time-consuming to acquire and analyze XRD data to recognize the crystal structure for each material\cite{oviedo2019fast}.
There are other machine learning models, different from XRD based methods, applied for space groups and crystal systems classification. Liu et al.\cite{liu2019using} trained a convolutional neural network model by atomic pair distribution function(PDF) for 45 most heavily represented space groups with an accuracy of 0.70. 
Kaufmann et al.\cite{kaufmann2020crystal} used a machine learning–based approach and developed a general methodology for rapid and autonomous identification of the crystal symmetry from electron backscatter diffraction(EBSD) patterns. However, those methods are inconvenient for predicting thousands of space groups since the input pictures of materials should be provided.
Theoretically, given the chemical composition of a material, computational prediction of its crystal structure is possible\cite{oganov2006crystal}. Several studies determine crystal structures by combining global optimization with DFT calculations\cite{oganov2006crystal}. These methods have demonstrated successes in a variety of cases. However, the DFT based methods generally require thousands of CPU hours and can only be applied to predict structures of relative small systems\cite{ziletti2018insightful}, which is not suitable for large-scale material space group determination.

Recently, there emerged several crystal structure prediction methods that start with a seed structure generated by the symmetry-restricted procedure \cite{sato2020adjusting,hu2021alphacrystal,lee2021crystal}. In these algorithms, usually for a given composition, a space group is specified to generate some random structures that satisfy the symmetry constraints of the space group. There is thus a need to predict the space group for a given composition. Several machine learning algorithms have been proposed for material crystal systems and space groups prediction quickly using composition information alone\cite{zhao2020machine,liang2020cryspnet}. 
In order to get the best classification performance, Zhao et al.\cite{zhao2020machine} uses two machine learning algorithms, random forest and multiple layer perceptron neural network models, combined with three kinds of descriptors, atom vector, one-hot encoding and Magpie, for the crystal system and space group classification. For the Material Project database they used, there are only 18 space groups selected, each has more than 1000 samples for multi-class classification which achieve a performance of 0.652 and 0.637 in terms of the F1-scores.
Liang et al.\cite{liang2020cryspnet} proposed CryspNet for crystal lattice type, space group and lattice constants prediction using deep neural networks. However, the classification accuracy scores only range from 0.638 and 0.907 in the fourteen crystal categories they divided.

In this work, we present an improved machine learning model for the crystal system and space group prediction by inputting the formulas of the crystal materials. Magpie descriptors are used as the basic descriptor set which is combined with a new descriptor set that we proposed, to train our machine learning models. To ensure that our classification algorithm can predict all types of crystal materials, we trained and validated the models on all kinds of entries from the Material Project Database of September 2020. 
 
Compared with previous works based only on the Magpie descriptors, the addition of the new descriptors enables our models to achieve significantly improved performance on crystal system and space group prediction. For example, the space group prediction of the cubic crystal system, which consists of 18,325 materials, has an accuracy score of 0.961. Our models can simultaneously make crystal system and space group predictions for a large number of hypothetical materials, which can make contribution to our knowledge-rich approach for crystal structure prediction. Our algorithm is also useful for downstream tasks such as the exploration of the structure and properties of new materials.

Our contributions can be summarized as follows:
\begin{itemize}
\item We propose a new descriptor set for crystal system and space group prediction of crystal materials which achieves significant performance improvements compared to prior studies. We also identified what features contribute most in our experiments for the classification performance
\item We remove the duplication caused by the isomer, and build ML models for multi-class classification for space group and crystal system of crystal materials.
\item We build ML models for multi-label classification for space group prediction and crystal system prediction.
\item We conduct extensive experiments with different machine learning algorithms. Our experiments show that our algorithm based on random forest achieves high performance in crystal system and space group prediction.

\end{itemize}
\FloatBarrier

\section{Materials and Methods}

\subsection{Datasets}

\begin{table}[h]
\begin{center}
\caption{ lattice parameter relationships for materials of different lattice systems. }
\label{latticepattern}
\begin{tabular}{|l|l|l|l|l|}
\hline
Crystal system & Edge lengths                                          & Axial angles                                    & Space groups &Amount \\ \hline
Cubic          & $a=b=c$                                                 & $\alpha=\beta=\gamma=90$                             & 195-230 & 18324    \\ \hline
Hexagonal      & $a=b$                                                   & $\alpha=\beta=90, \gamma=120$                        & 168-194 & 9243     \\ \hline
Trigonal       & $a=b\neq c$                         & $\alpha=\beta=90, \gamma=120$                        & 143-167 & 11086      \\ \hline
Tetragonal     & $a=b\neq c$                         & $\alpha=\beta=\gamma=90$                             & 75-142 & 14654      \\ \hline
Orthorhombic   & $a\neq b\neq c$ & $\alpha=\beta=\gamma=90$                             & 16-74 & 26800      \\ \hline
Monoclinic     & $a\neq c$ & $\alpha=\gamma=90, \beta\neq 90$ & 3-15 &29872        \\ \hline
Triclinic      & all other cases                                       & all other cases                                 & 1-2 &15297         \\ \hline
\end{tabular}
\end{center}
\end{table}

There are 125,276 inorganic material items used in our experiment, which are extracted from the Materials Project\cite{jain2013commentary}, an extensive material database which includes the properties of all known inorganic materials. According to the degree of geometric forms symmetry, those crystals are divided into seven categories, namely cubic crystal system, hexagonal crystal system, tetragonal crystal system, trigonal crystal system, orthorhombic crystal system, monoclinic crystal system and triclinic crystal system. The amount of those crystal systems and some details of geometric forms symmetry are shown in Table \ref{latticepattern}. 
For those crystals, they are divided into 230 different combinations of symmetrical elements. Each space group has a unique crystal system corresponding to it. 

\begin{figure}[h]
	\centering
	\begin{subfigure}{.45\textwidth}
		\includegraphics[width=\textwidth]{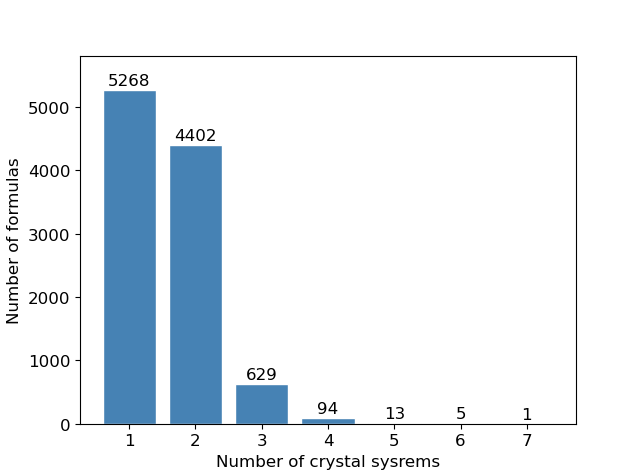}
		\caption{The relationship of molecular formula quantity and crystal system quantity.}
		\vspace{3pt}
	\end{subfigure}
	\begin{subfigure}{.45\textwidth}
		\includegraphics[width=\textwidth]{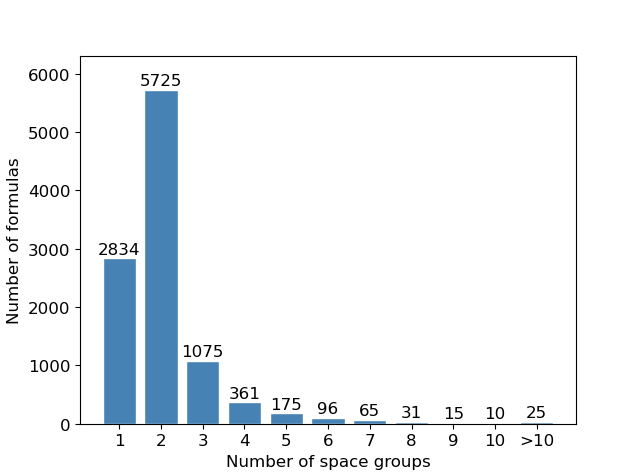}
		\caption{The relationship of molecular formula quantity and space group quantity.}
		\vspace{3pt}
	\end{subfigure}
	\caption{The statistics of the formulas which have isomers.}
	\label{fig:distribution}
\end{figure}

However, for a given crystal composition, there may be multiple structural isomers which share the same chemical formula. There are a total of 10,412 unrepeated formulas that have isomers in the dataset we used. As is shown in Figure\ref{fig:distribution}, we show several statistics  of these 10,412 formulas. We can see that there are some formulas that have only one crystal system and even one space group even though they have isomers. This is because the isomers of above formulas belong to the same crystal system or space group. In addition, there are no more than two crystal systems for most materials formula isomers, but there is a formula that has even 7 crystal systems. For material formulas with isomers, more than half of the formulas have two space groups. In order to make each formula match only one crystal system and space group, we calculate the retain score by formula\ref{con:retain_score}, which is also used in cryspnet\cite{liang2020cryspnet}, and keep the material structure with the highest score among the isomers while dropping out other materials. Then, our machine learning model has a unique crystal system or space group target for a formula to achieve multi-class classification. Moreover, we also trained machine learning models for multi-label classification \cite{zhao2020machine} for the 10,412 crystals which have isomers.

\begin{equation}
Score\left(f,s\right)=\frac{Abundance\left(f,s\right)}{E_{hull}\left(f,s\right)+\alpha}
\label{con:retain_score}
\end{equation}

In this formula, $\alpha$ is tunable for balancing the formation energy term and the abundance count from the Materials Project dataset, we set it as 0.1 in our work, $s$ and $f$ represent the space group and chemical formula. The $Abundance(s,f)$ means the number of records which have the same formula and space group. The $E_{hull}(s,f)$ is used to find the lowest formation energy above the convex hull by the given composition and space group. For the formulas that have isomers, we choose the material with the highest $Score(s,f)$ for our benchmark experiments. Then, there are 102,528 materials used in our experiment for multiclass classification.

\FloatBarrier
\subsection{Descriptors}

Magpie descriptors\cite{ward2016general} are a set of composition based materials attributes that calculate the statistics of stoichiometric attributes, elemental properties, electronic structure attributes and ionic compound attributes. It has been widely used for building machine learning models for composition based materials property predictions. In our work, the elemental property statistics have been used as the baseline data set for space group and crystal system prediction. There are 22 kinds of features in Magpie element property statistics including Atomic Number, Mendeleev Number, Atomic Weight, Melting Temperature, Periodic Table Row and Column, Covalent Radius, Electronegativity, the number of Valence e in each Orbital(s, p. d, f, total), the number of unilled er in each orbital (s, p. d, f, total),
Ground State Volume, Ground State Band Gap Energy, Ground State Magnetic Moment, and the Space Group Number of elements. The main features of the Magpie feature set are obtained by calculating the mean, average deviation, range, mode, minimum, and maximum of above elemental properties (weighted by the fraction of each element in the composition) to transform raw materials data into a form compatible with machine learning. We also add some other Magpie properties, which are used in cryspnet\cite{liang2020cryspnet} for structure information prediction, include Stoichiometry p-norm (p=0,2,3,5,7), Elemental Fraction, Fraction of Electrons in each Orbital, Band Center, Ion Property (possible to form ionic compound, ionic charge) to improve classification performance.

\begin{table}[]
\begin{center}
\caption{Descriptors.}
\label{descriptors}
\begin{tabular}{|c|c|c|}
\hline
Element Property statistics of Magpie & Additional Predictors of Magpie & Added Predictors in this work  \\ \hline
Atomic Number & Stoichiometry p-norm (p=0,2,3,5.7) & Total Atom Number \\ \hline
Mendeleev Number & Elemental Fraction & Maximum Atom Number \\ \hline
Atomic Weight & Fraction of Electrons in each Orbital & Minimum Atom Number \\ \hline
Melting Temperature & Band Center & Average Atom Number \\ \hline
Periodic Table Row and Column & \begin{tabular}[c]{@{}c@{}}lon Property (possible to form\\ ionic compound, ionic charge)\end{tabular} & Specific Value \\ \hline
Covalent Radius &  & Atom Number Variance \\ \hline
Electronegativity &  &  \\ \hline
\begin{tabular}[c]{@{}c@{}}The number of Valence e in\\ each Orbital(s, p. d, f, total)\end{tabular} &  &  \\ \hline
\begin{tabular}[c]{@{}c@{}}The number of unilled e in\\ each orbital (s, p. d, f, total)\end{tabular} &  &  \\ \hline
Ground State Volume &  &  \\ \hline
Ground State Band Gap Energy &  &  \\ \hline
Ground State Magnetic Moment &  &  \\ \hline
Space Group Number &  &  \\ \hline
\end{tabular}
\end{center}
\end{table}

Additionally, we propose a set of new descriptors
to improve crystal systems and space groups prediction performance. Our descriptors are related to the number of various atoms in a crystal structure. For a crystal, the numbers of atoms is different for different elements within a unit cell and our new descriptor is obtained by (for an element) calculating the maximum atom number, the minimum atom number, the total atom number, the average atom number, the specific value (ratio of pretty formula to full formula) and the atom number variance of all the elements within a crystal structure. All the features used in our experiments are shown in table\ref{descriptors}


\subsection{ Machine learning models: Random forest(RF), Extreme Gradient Boosting(XGBoost) and Deep Neural Networks(DNN)}
In this study, we use ensemble machine learning algorithms and neural networks to find out the best model for crystal system and space group prediction.

\subsubsection{Random Forest}
Random forest (RF)\cite{breiman2001random}, a popular ensemble machine learning algorithm, is one of the major bagging machine learning models. Its driving principle in classification is to build several estimators independently and each estimator gets the probability of possible output labels. Then the average of the predicted probabilities of all estimators are calculated as outputs. Labels with the highest average probability are used as the output results. In our random forest classification models, we choose 'entropy' as our criterion. The two important hyper-parameters, the number of trees and max features, are set to be 100 and 80 respectively. The max depth is None. The min samples leaf and min samples split are sat as 1 and 2. This algorithm was implemented by the Scikit-Learn\cite{pedregosa2011scikit} library in Python 3.6

\subsubsection{Extreme Gradient Boosting}
XGBoost\cite{chen2016xgboost} is an excellent boosting algorithm which is widely used in machine learning classification tasks. As an integrated algorithm, xgboost has outstanding performance in classification with excellent generalization performance. In addition, it is optimized by a series of methods, such as parallel processing, supporting regularization and using second-order Taylor expansion for loss function. Similar to traditional boosting algorithms which are composed of several weak algorithms, when training the XGBoost model, each weak algorithm tries to correct the error of the previous algorithms. In our work, we chose the 'gbtree' as our booster whose max depth was set as 6. As an important parameter, the number of the trees is set as 180. The learning rate, alpha, gamma, lambda are set as 0.3, 0, 0 ,1 respectively. For triclinic, which has only two kinds of space groups, the objective (used to specify the learning task and the corresponding learning objective) and evaluation metrics are set as softmax and multiclass logloss. For the rest experiments of the XGBoost model, we choose the logistic and logloss as the objective and evaluation metrics. We use the XGBoost library in Python 3.6 to implement this algorithm.

\subsubsection{Deep Neural Networks}

\begin{figure}[ht]
  \centering
  \includegraphics[width=0.6\linewidth]{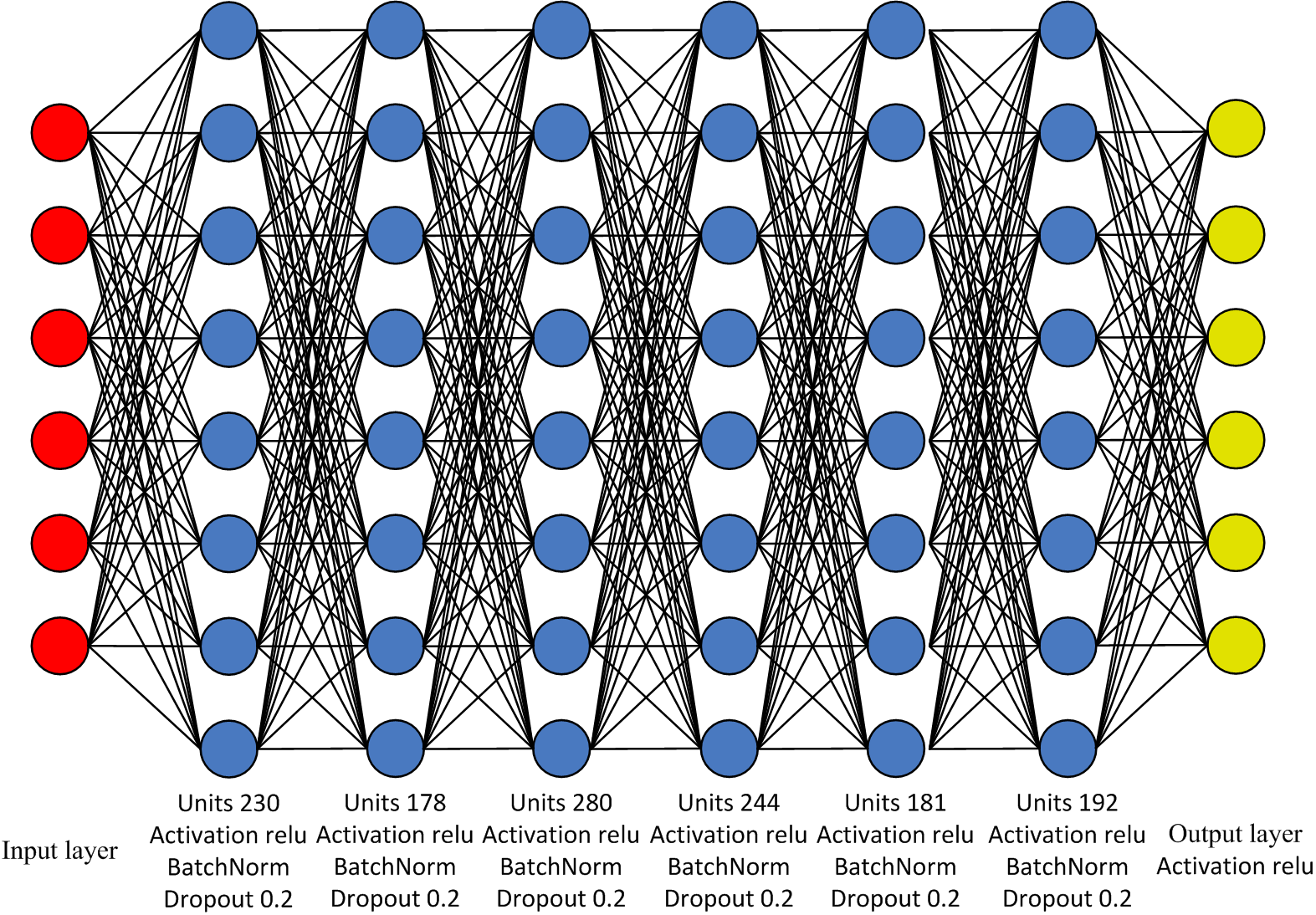}
  \caption{Architecture of the deep neural network.}
  \label{fig:NN}
\end{figure}

The deep neural networks, which plays an important role in material performance prediction and discovery of new materials, is also used in our work. The structure of our NN model, shown in Figure\ref{fig:NN}, is composed of 7 fully connected layers, and the numbers of the hidden nodes are 230, 178, 280, 244, 181, 192 respectively. Relu\cite{krizhevsky2012imagenet} is used as the action function for those layers. And after each layer except the last one, two strategies, Dropout and BatchNorm, are used to prevent overfitting. The loss function is set as the categorical cross entropy loss. The Adam optimizer is used in model training. The learning rate, epochs and batch size are set as 0.001, 2000 and 255 respectively. In order to prevent overfitting and improve  performance of the NN, we use the early stopping strategy where the monitor is 'loss' and patience is 30. Our NN model is implemented on TensorFlow2.4.

\subsection{Evaluation criteria}

We use K-fold cross-validation in our work to assess the performance of classification models of different algorithms. The process of K-fold cross-validation strategy is that it randomly splits the initial sample set into K sub-sample sets, taking one of them as the test set, the rest as the training set. After the initial sample set splitted into K sub-sample sets, the cross-validation is repeated K times, each sub-sample set is verified once, then the results of K times are averaged to finally obtain a single performance estimate. The 10-fold cross-validation method is chosen to evaluate the classification performance of different algorithms in our work. Due to the uneven distribution of samples, the following performance criteria in this work are used, including accuracy, Matthews correlation coefficient(MCC), weighted precision, weighted recall and weighted F1 score\cite{murphy2012machine,tao2021machine}.

\FloatBarrier
\section{Results and Discussion}
\subsection{Multi-class classification}
\subsubsection{Prediction performance of multi-class classification}

\begin{table}[h]
\begin{center}
\caption{The classification performance of RF algorithm for all data removed the duplication caused by isomers.}
\label{all_data_classification}
\begin{tabular}{|c|l|l|l|l|l|}
\hline
\multicolumn{1}{|l|}{} & \multicolumn{1}{c|}{Accuracy} & \multicolumn{1}{c|}{MCC} & \multicolumn{1}{c|}{Precision} & \multicolumn{1}{c|}{Recall} & \multicolumn{1}{c|}{F1 score} \\ \hline
Crystal system prediction & 0.816$\pm$0.005 & 0.779$\pm$0.006 & 0.818$\pm$0.005 & 0.816$\pm$0.005 & 0.816$\pm$0.005 \\ \hline
Space group prediction & 0.729$\pm$0.004 & 0.721$\pm$0.004 & 0.734$\pm$0.004 & 0.729$\pm$0.004 & 0.725$\pm$0.004 \\ \hline
\end{tabular}
\end{center}
\end{table}

\begin{figure}[ht]
  \centering
  \includegraphics[width=0.45\linewidth]{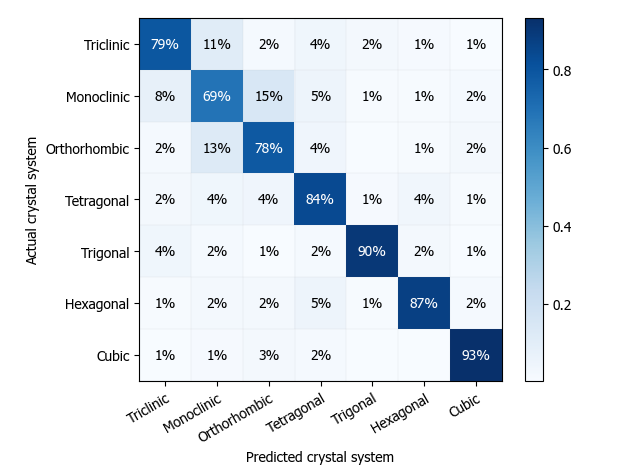}
  \caption{Confusion matrix of crystal system classification.}
  \label{fig:crystal_system}
\end{figure}

We trained the multi-class prediction models using the data sets after removing the duplications caused by isomers. Then we use the 10-fold cross-validation evaluation approach to assess our model performance. 
The crystal system and space group classification performance for all data are shown in Table \ref{all_data_classification}. 
Our random forest model achieves an accuracy score of 0.816 and F1 score of 0.812 for crystal systems prediction, and its confusion matrix is shown in Figure \ref{fig:crystal_system} which shows that the more regular of the crystal material structures, the better classification result the model has in general. For space groups prediction with around 230 categories, the accuracy and MCC scores are 0.729 and 0,721 by just using the formula as the input to the machine learning models.

\begin{table}[h]
\begin{center}
\caption{The performance of space group classification in different crystal systems.}
\label{space_group_classification}
\begin{tabular}{|c|l|l|l|l|l|l|}
\hline
\multicolumn{1}{|l|}{} & \multicolumn{1}{c|}{data set size} & \multicolumn{1}{c|}{Accuracy} & \multicolumn{1}{c|}{MCC} & \multicolumn{1}{c|}{Precision} & \multicolumn{1}{c|}{Recall} & \multicolumn{1}{c|}{F1 score} \\ \hline
Cubic & 17367 & 0.961$\pm$0.006 & 0.945$\pm$0.008 & 0.960$\pm$0.005 & 0.961$\pm$0.006 & 0.959$\pm$0.006 \\ \hline
Hexagonal & 8201  & 0.909$\pm$0.008 & 0.888$\pm$0.010 & 0.908$\pm$0.008 & 0.909$\pm$0.008 & 0.906$\pm$0.008 \\ \hline
Trigonal &  9429 & 0.824$\pm$0.012 & 0.797$\pm$0.014 & 0.823$\pm$0.013 & 0.824$\pm$0.012 & 0.818$\pm$0.012 \\ \hline
Tetragonal &   12675  & 0.849$\pm$0.013 & 0.832$\pm$0.015 & 0.846$\pm$0.013 & 0.849$\pm$0.013 & 0.840$\pm$0.014 \\ \hline
Orthorhombic &   22392 & 0.755$\pm$0.005 & 0.729$\pm$0.006 & 0.759$\pm$0.005 & 0.755$\pm$0.005 & 0.746$\pm$0.006 \\ \hline
Monoclinic &   23024  & 0.712$\pm$0.009 & 0.647$\pm$0.011 & 0.715$\pm$0.010 & 0.712$\pm$0.009 & 0.703$\pm$0.010 \\ \hline
Triclinic &  9440  & 0.835$\pm$0.013 & 0.665$\pm$0.026 & 0.835$\pm$0.013 & 0.835$\pm$0.013 & 0.834$\pm$0.013 \\ \hline
All & 102528 & 0.729$\pm$0.004 & 0.721$\pm$0.004 & 0.734$\pm$0.004 & 0.729$\pm$0.004 & 0.725$\pm$0.004 \\ \hline
\end{tabular}
\end{center}
\end{table}

The results of space group classification in each crystal system are shown in Table \ref{space_group_classification}, the accuracy score of our RF model reaches 0.961 in space group classification for cubic materials. Although the crystal structures in the cubic crystal system have the highest symmetry compared to other crystal systems, the Cubic crystal system has 36 space groups, from the space group of 195 to 230. It is exciting to achieve such high space group classification performance with so many categories. However, the classification performances in Orthorhombic and Monoclinic for space group prediction are much lower with accuracy scores neither reaching 0.8. This is because these two crystal systems have complex crystal structures, and the input of our machine learning models is only the crystal formula. It is difficult to get the complex crystal structures information merely by the formulas of materials, which can however be ameliorated with large dataset. While the top-1 accuracy scores may not be ideal, it is possible to use the top-k prediction results of our models in downstream tasks to improve the hit rate. For other crystal systems, the performances in space groups classification have scores above 0.80 in terms of accuracy, MCC and  F1 scores.

\begin{figure}[h]
	\centering
	\begin{subfigure}{.9\textwidth}
		\includegraphics[width=0.80\linewidth]{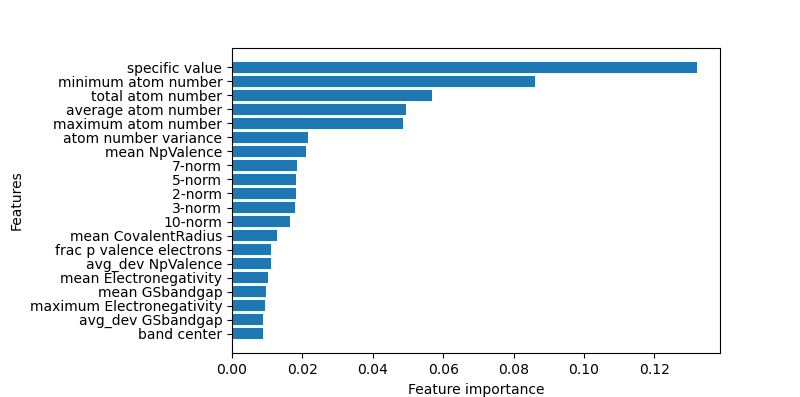}
		\caption{feature importance of crystal system classification.}
		\vspace{3pt}
	\end{subfigure}
	\centering
	\begin{subfigure}{.9\textwidth}
		\includegraphics[width=0.80\linewidth]{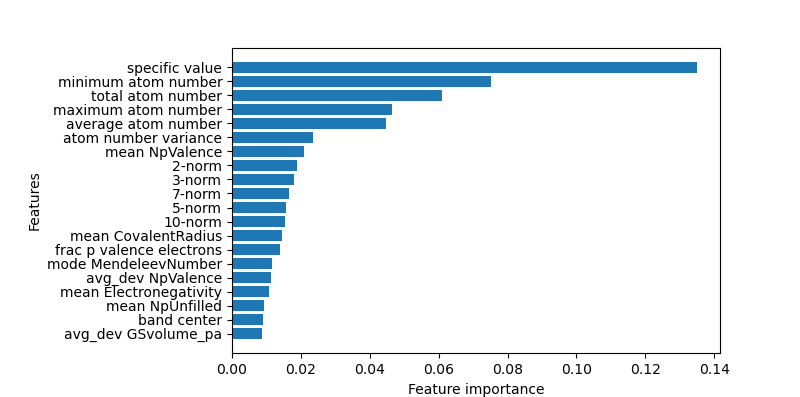}
		\caption{feature importance of space group classification.}
		\vspace{3pt}
	\end{subfigure}
	\caption{Feature importance of crystal system and space group classification.}
	\label{fig:feature_importance}
\end{figure}

In order to find out which features most affect the performance of classification, we calculated the feature importance scores to sort the top 20 features which are shown in Figure \ref{fig:feature_importance}. We find that the specific value, minimum atom number,  total atom number, average atom number, maximum atom number and atom number variance have major impact, which explains why our model is better than previous Magpie features based machine learning methods.

\subsubsection{Performance comparison with previous works}

\begin{figure}[h]
	\centering
	\begin{subfigure}{.7\textwidth}
		\includegraphics[width=\textwidth]{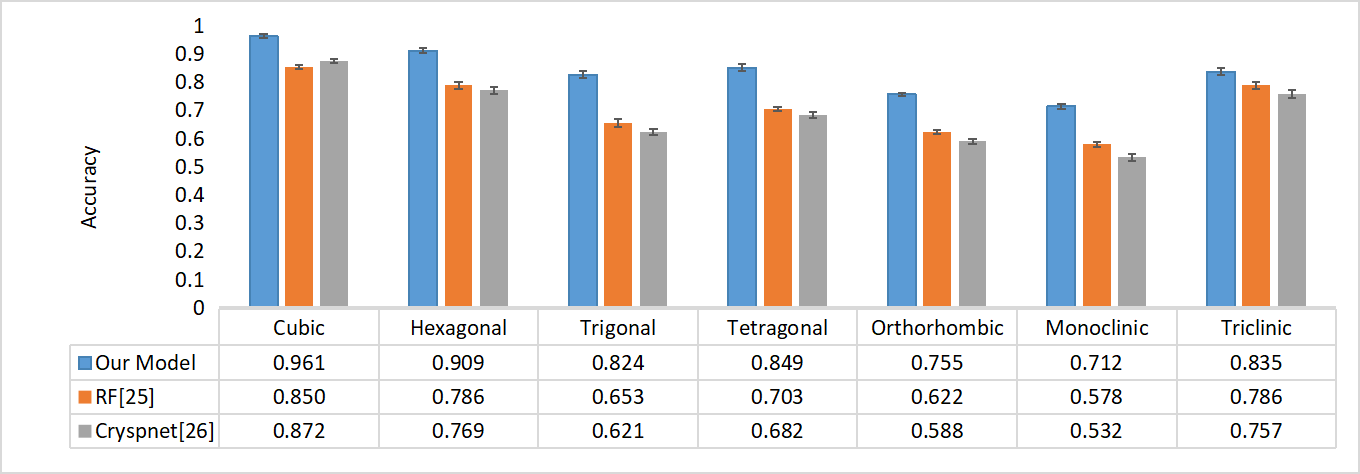}
		\caption{Performance comparison in terms of Accuracy}
		\vspace{3pt}
	\end{subfigure}
	\begin{subfigure}{.7\textwidth}
		\includegraphics[width=\textwidth]{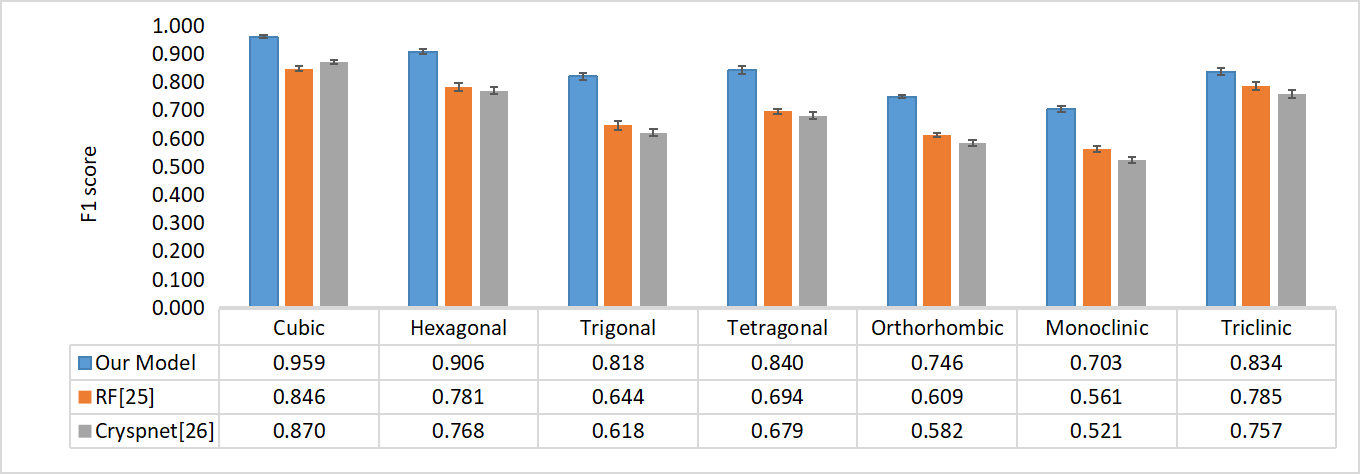}
		\caption{Performance comparison in terms of F1 score}
		\vspace{3pt}
	\end{subfigure}
	\caption{
	Performance comparison with previous works for space group prediction. All those performances are trained by the same dataset. Note that the results in RF\cite{zhao2020machine} and Cryspnet\cite{liang2020cryspnet} are based on our implementations of the corresponding algorithms described in the literature.
	}
	\label{fig:compare_dataset}
\end{figure}

\begin{figure}[h]
	\centering
	\begin{subfigure}{.45\textwidth}
		\includegraphics[width=\textwidth]{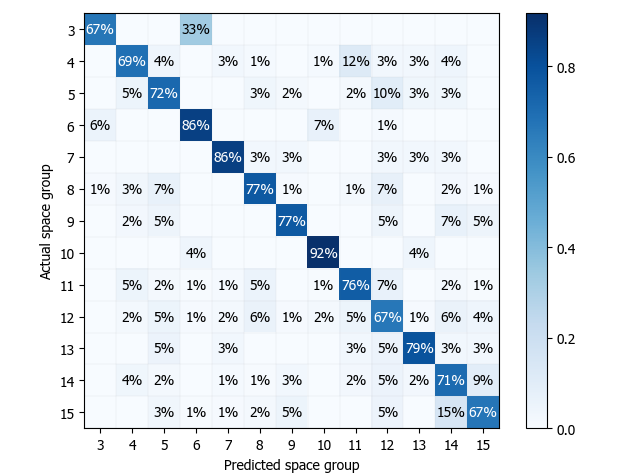}
		\caption{RF model with all descriptors in Table\ref{descriptors}.}
		\vspace{3pt}
	\end{subfigure}
	\begin{subfigure}{.45\textwidth}
		\includegraphics[width=\textwidth]{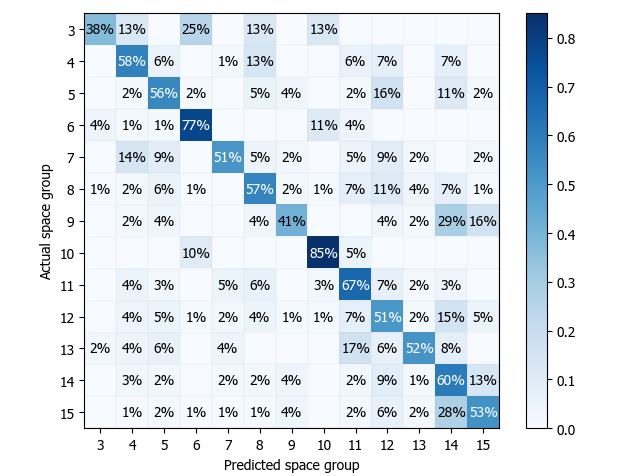}
		\caption{RF model with only the new descriptor set we proposed.}
		\vspace{3pt}
	\end{subfigure}
	\begin{subfigure}{.45\textwidth}
		\includegraphics[width=\textwidth]{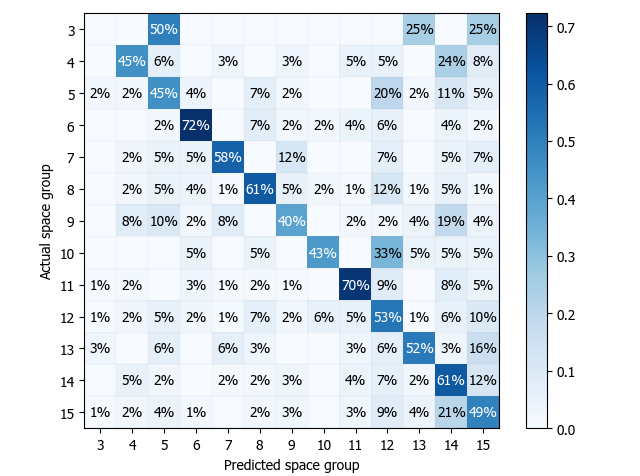}
		\caption{RF model in reference\cite{zhao2020machine}.}
		\vspace{3pt}
	\end{subfigure}
	\begin{subfigure}{.45\textwidth}
		\includegraphics[width=\textwidth]{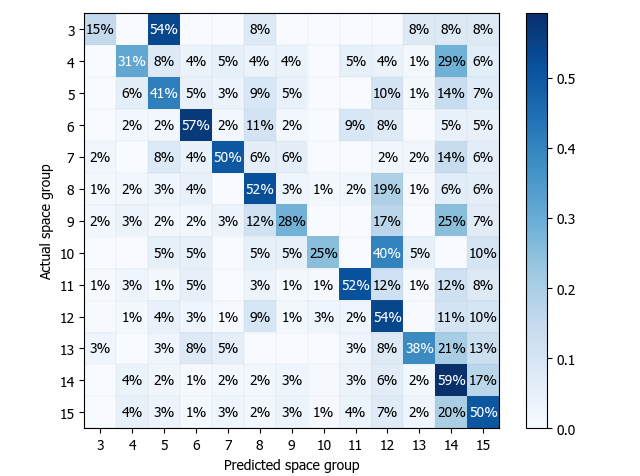}
		\caption{NN of cryspnet\cite{liang2020cryspnet}.}
		\vspace{3pt}
	\end{subfigure}
	
	\caption{Performance of space group prediction for monoclinic materials using different algorithms. All those performances are trained by the same dataset. Note that the results in (c) and (d) are based on our implementations of the corresponding algorithms described in the literature.}
	\label{fig:performance_mono_mp}
\end{figure}

For multi-class classification, we select all kinds of space groups available in Materials Project in order that our classification algorithm can predict all types of crystal materials. However, in ref.\cite{zhao2020machine}, there are only 18 space groups selected for experiment, each having more than 1000 compositions for space group prediction. In addition, we use Formula \ref{con:retain_score} to tackle the duplicate formula caused by isomers, but in paper\cite{zhao2020machine}, there is no information about the isomer processing. 
In Magpie descriptors based space group classification, compared with Cryspnet\cite{liang2020cryspnet}, we use random forest algorithm which has better performance than the neural network\cite{zhao2020machine}. 
And most importantly, we propose a new descriptor set, which greatly improves the performance of our model compared to previous works on space group prediction for inorganic materials\cite{zhao2020machine,liang2020cryspnet}.
As is shown in Figure \ref{fig:compare_dataset}, we make comparisons with two previous works for space group prediction.
Our model is the random forest in this work, which is trained using all the descriptors in Table\ref{descriptors}.
RF\cite{zhao2020machine} is the random forest algorithm trained by Element Property statistic descriptors of Magpie. Cryspnet\cite{liang2020cryspnet} is the neural networks framework of cryspnet trained by Element Property statistic descriptors and additional Predictors of Magpie. 
The three models are trained using the same dataset of materials project processed by Formula \ref{con:retain_score}.
As we can see that compared with the previous works of cryspnet and ref.\cite{zhao2020machine}, the prediction performance in terms of accuracy scores are improved by about 0.14 and 0.16 respectively in the six crystal systems except the triclinic. For F1 scores, the improvements are similar. Moreover, the higher the symmetry of the crystal structures, the better prediction performance our RF can achieve for space group classification in the three algorithms. 

In particular, we checked the space group prediction performance in the Monoclinic system which has 13 space groups and relatively complex crystal structures to explore some classification details. From the confusion matrices shown in Figure \ref{fig:performance_mono_mp}, it can be found that the performance has obvious improvements in crystal systems classification by using the descriptor we proposed. The improvement of the space group number of 3 and 10 are obvious. To illustrate the validity of our descriptor, we make classification by the combination of random forest algorithm and the new descriptors we proposed, the results are shown in Figure \ref{fig:performance_mono_mp}(b). The model even has a better performance than the two previous works. This shows that the descriptors about the number of atoms in the lattice are effective for space group classification.

\subsubsection{Performance comparison of different algorithms}

\begin{figure}[h]
	\centering
	\begin{subfigure}{.7\textwidth}
		\includegraphics[width=\textwidth]{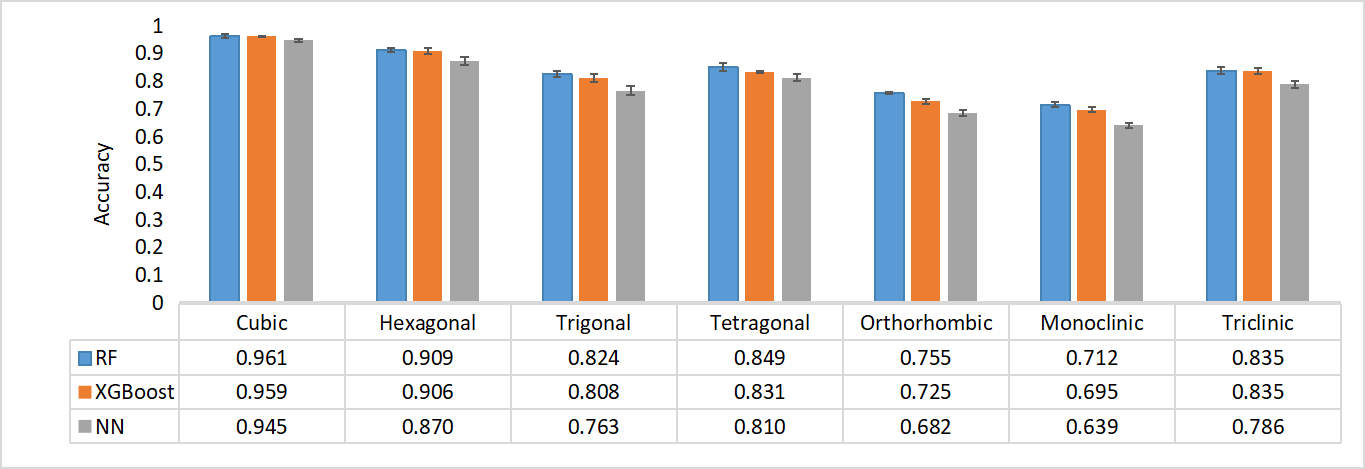}
		\caption{Performance comparison in terms of Accuracy}
		\vspace{3pt}
	\end{subfigure}
	\begin{subfigure}{.7\textwidth}
		\includegraphics[width=\textwidth]{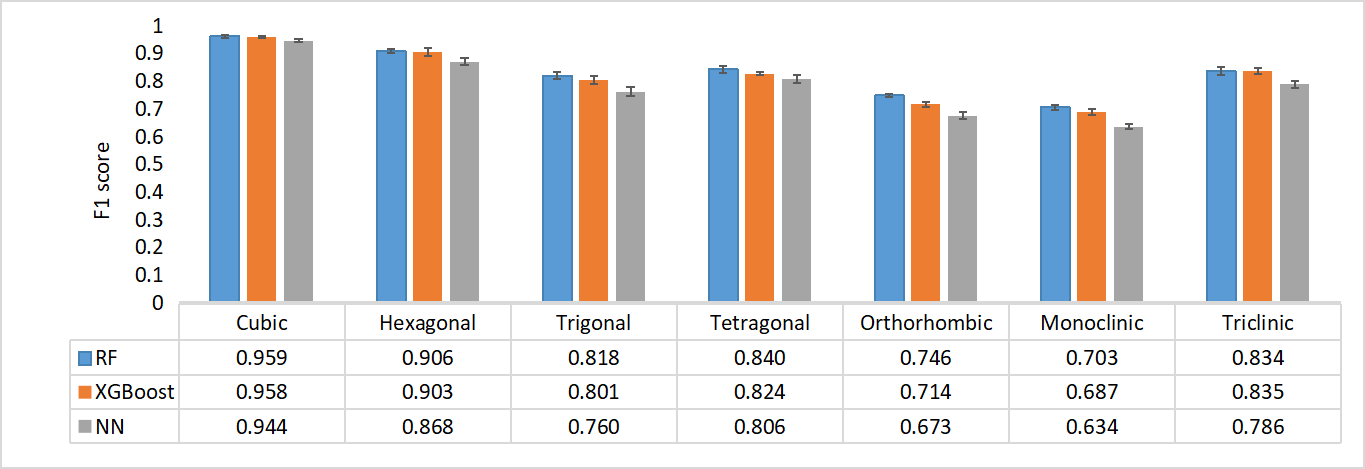}
		\caption{Performance comparison in terms of F1 score}
		\vspace{3pt}
	\end{subfigure}
	\caption{
	Performance comparison of different algorithms in our work for space group prediction}
	\label{fig:per_dif_alg}
\end{figure}

Various machine learning algorithms have been used for structural information prediction. Here, we evaluate the performance of three powerful machine learning methods, the DNN, RF, and XGBoost, in space group and crystal system classification.
As is shown in Figure \ref{fig:per_dif_alg}, the RF model performances in both accuracy and F1 score are slightly better than those of XGboost. And in our experiment, which uses the physical and chemical properties as descriptors, the performance of the two kinds of ensemble tree algorithms is better than that of NN. In this work, RF has demonstrated better performance than NN in making crystal systems and space groups classification, which is consistent with the results in previous study\cite{zhao2020machine}. However, there are many works indicating that NN has excellent performance compared with other machine learning models in the field of material informatics\cite{oviedo2019fast,suzuki2020symmetry}. Therefore, for different problems, choosing suitable machine learning algorithms is needed to achieve the best results.

\subsection{Multi-label classification}

Considering the fact that some formulas can correspond to multiple structures of different crystal systems or space groups, the prediction problem of crystal systems and space groups can also be mapped as a multi-label classifiation problem\cite{zhao2020machine}. Here we train machine learning models for multi-label classification for the 10,412 crystals which have isomers. The crystal system classification performance is shown in Tabel \ref{multi_lable} which has the recall score of 0.706, the precision score of 0.813, and the F1 score is 0.751. For space group classification, the performance is inferior to the crystal system prediction. The precision, recall and F1 scores are 0.764, 0.452 and 0.547 respectively. This is because the crystal structures has 230 space groups, which are far more diverse than the seven types of crystal systems.

\begin{table}[]
\begin{center}
\caption{The performance of multi-label classification.}
\label{multi_lable}
\begin{tabular}{|c|l|l|l|}
\hline
\multicolumn{1}{|l|}{} & \multicolumn{1}{c|}{Precision} & \multicolumn{1}{c|}{Recall} & \multicolumn{1}{c|}{F1 score} \\ \hline
Crystal system prediction & 0.813$\pm$0.011 & 0.706$\pm$0.013 & 0.751$\pm$0.010 \\ \hline
Space group prediction & 0.764$\pm$0.020 & 0.452$\pm$0.018 & 0.547$\pm$0.017 \\ \hline
\end{tabular}
\end{center}
\end{table}
\FloatBarrier

\section{Conclusion}
Computational prediction of space groups and crystal systems plays an important role in analyzing crystal material structures and their physical and chemical properties and is useful for crystal structure prediction. While there are various methods to classify the space groups and crystal systems of crystal materials in previous works, this study proposes an efficient and easy-to-use method to achieve more accurate classification by introducing a new set of materials composition based descriptors. When combined with the Magpie descriptors, our algorithms' classification performance have improved significantly compard to previous algorithms that use only the magpie descriptors. Our trained models and source code are freely available at \url{https://github.com/Yuxinya/SG_predict}, which should be helpful for downstream works such as material property exploration, material screening, and crystal structure prediction.

\section{Contributions}

Conceptualization, J.H. and Y.L.; methodology, Y.L. and J.H.; software, Y.L. and J.H.; validation, Y.L. and J.H.;investigation, Y.L.,J.H., R.D., and W.Y.; resources, J.H.; writing–original draft preparation, J.H. and Y.L.; writing–review and editing, J.H and Y.L.; visualization, Y.L., R.D; supervision, J.H.; funding acquisition, J.H

\section{Data Availability}
The data that support the findings of this study are openly available in Materials Project database\cite{jain2013commentary} at \url{http://www.materialsproject.org}.

\section{Acknowledgement}
Research reported in this work was supported in part by NSF under grant and 1940099 and 1905775. The views, perspective,and content do not necessarily represent the official views of the NSF

\bibliographystyle{unsrt}  
\bibliography{references}  

\end{document}